\documentclass{aa} 
\usepackage{graphicx}
\usepackage{txfonts}
\usepackage{natbib}
\bibpunct{(}{)}{;}{a}{}{,}

\begin{document}
   \title{Mid-infrared interferometry of massive young stellar objects}

   \subtitle{I. VLTI and Subaru observations of the enigmatic object M8E-IR\thanks{
             Based on observations within the ESO programs 073.C-0175(A),
	     273.C-5044(A), 074.C-0389(B), and 075.C-0755(A,B). 
	     }}
   \author{H.~Linz\inst{1}\and
          Th.~Henning\inst{1}\and
	   M.~Feldt\inst{1}\and
	   I.~Pascucci\inst{2}\and
	   R.~van Boekel\inst{1}\and
	   A.~Men'shchikov\inst{3}\and
	   B.~Stecklum\inst{4}\and
	   O.~Chesneau\inst{5}\and
	   Th. Ratzka\inst{6}\and
	   S.~P.~Quanz\inst{1}\and
	   Ch.~Leinert\inst{1}\and
	   L.B.F.M.~Waters\inst{7}\and
	   H.~Zinnecker\inst{6}
          }

   \offprints{H.~Linz}

   \institute{Max-Planck-Institut f\"ur Astronomie, 
              K\"onigstuhl 17, D-69117 Heidelberg, Germany\\
              \email{[linz,henning,feldt,leinert,boekel,quanz]@mpia-hd.mpg.de}
         \and
              Department of Physics and Astronomy, Johns Hopkins 
	      University, 403 Bloomberg Center, 3400 N.~Charles Street,
	      Baltimore, \\ MD 21218, 
             \email{pascucci@pha.jhu.edu}
	 \and
	      CEA, IRFU, SAp, Centre de Saclay, F-91191 Gif-sur-Yvette, France, 
	      \email{alexander.menshchikov@cea.fr}
	 \and
	      Th\"uringer Landessternwarte Tautenburg, Sternwarte 5,
	      D--07778 Tautenburg, Germany,
	      \email{stecklum@tls-tautenburg.de}
	 \and
	      UMR 6525, Univ.~Nice, CNRS, Obs.~de
              la C\^{o}te d'Azur, Av.~Copernic, F-06130 Grasse, France,
	      \email{chesneau@obs-azur.fr}
	 \and 
	      Astrophysikalisches Institut Potsdam,  
	      Sternwarte 16, D-14482 Potsdam, Germany, 
	      \email{[tratzka,hzinnecker]@aip.de}	      
	 \and
	      Astronomical Institute ``Anton Pannekoek'', Universiteit van  
	      Amsterdam, Kruislaan 403, NL-1098 SJ Amsterdam, Netherlands \\
	      \email{rensw@science.uva.nl}
             }

   \date{Accepted July 02, 2009}

 
  \abstract
   {Our knowledge of the inner structure of embedded massive young stellar 
    objects is still quite limited. Thus, it is difficult to decide to what
    extent the mass accumulation onto forming massive stars differs from
    the process of low-mass star formation.}
   {We attempt to overcome the spatial resolution limitations of conventional thermal 
    infrared imaging.}
   {We employed mid-infrared interferometry using the MIDI instrument on the
    ESO/VLTI facility  to investigate M8E-IR, a well-known massive young stellar object
    suspected of containing a circumstellar disk.  Spectrally dispersed visibilities in
    the 8--13 $\mu$m range have been obtained at seven interferometric 
    baselines.}
   {We resolve the mid-infrared emission of M8E-IR and find typical sizes of the emission
   regions of the order of 30 milliarcseconds ($\cor 45$ AU). Radiative transfer 
   simulations were performed to interpret the data. 
   The fitting of the spectral energy distribution,
   in combination with the measured visibilities, does not provide evidence for an extended circumstellar 
   disk with sizes $\ga 100$ AU but requires the presence of an extended envelope. 
   The data are not able to constrain the presence of a small-scale disk in addition to an envelope.
   In either case, the interferometry measurements indicate 
   the existence of a strongly bloated, relatively cool central object, possibly tracing the
   recent accretion history of M8E-IR. In addition, we 
   present 24.5\,$\mu$m images that clearly distinguish between 
   M8E-IR and the neighbouring ultracompact H{\sc ii} region and which show the 
   cometary-shaped infrared morphology of the latter source.}
   {Our results show that IR interferometry, combined with radiative transfer modelling, 
   can be a viable tool to reveal
   crucial structure information on embedded massive young stellar objects 
   and to resolve ambiguities arising from fitting the  SED.}


   \maketitle
%

\section{Introduction}

High-mass stars predominantly form in clustered environments much farther away
from the Sun, on average, than typical well-investigated low-mass star-forming 
regions. Thus, high spatial resolution is a prerequisite for making progress in 
the observational study of high-mass star formation. Furthermore, all phases 
prior to the main sequence are obscured by
dense circumstellar environments. This forces observers of deeply embedded massive
young stellar objects (MYSOs) to move to the mid-infrared (MIR), where the 
resolution of conventional imaging is limited to $>$ 0\farcs25 even with 8-m
class telescopes. Hence, one traces linear scales still several hundred AU in 
size even for the nearest MYSOs, and conclusions on the geometry of the 
innermost circumstellar material remain ambiguous. MIR emission moderately 
resolved with single-dish telescopes could even arise from the inner outflow 
cones \citep[e.g.,][]{2006ApJ...642L..57D,2005A&A...429..903L}. \\
A way to overcome the diffraction limit of single telescopes is to employ
interferometric techniques. We are presently conducting a larger survey toward 
MYSOs based on MIR interferometry. While the results for the other objects will 
be reported in subsequent publications, we concentrate here on the object 
\object{M8E-IR}, a prominent BN-type MYSO 
\citep[cf.][]{1990FCPh...14..321H} at a distance of 1.25--1.5 kpc
\citep{2007MNRAS.374.1253A,1984ApJ...278..170S}. 
Although M8E-IR had been well investigated in the 1980s, the spatial 
resolution for most of the IR observations of M8E-IR was poor. An
exception is the work by \citet{1985ApJ...298..328S} who speculated on the 
existence of a small circumstellar disk around M8E-IR based on thermal infrared
lunar occultation data. 
Here, we present our work on M8E-IR which includes 8--13 $\mu$m
interferometry to dissect the MYSO M8E-IR itself, and N- and Q-band 
imaging to have a fresh look at M8E-IR in its relation to the environment.
   \begin{figure*}
   \centering
   \includegraphics[width=7.75cm]{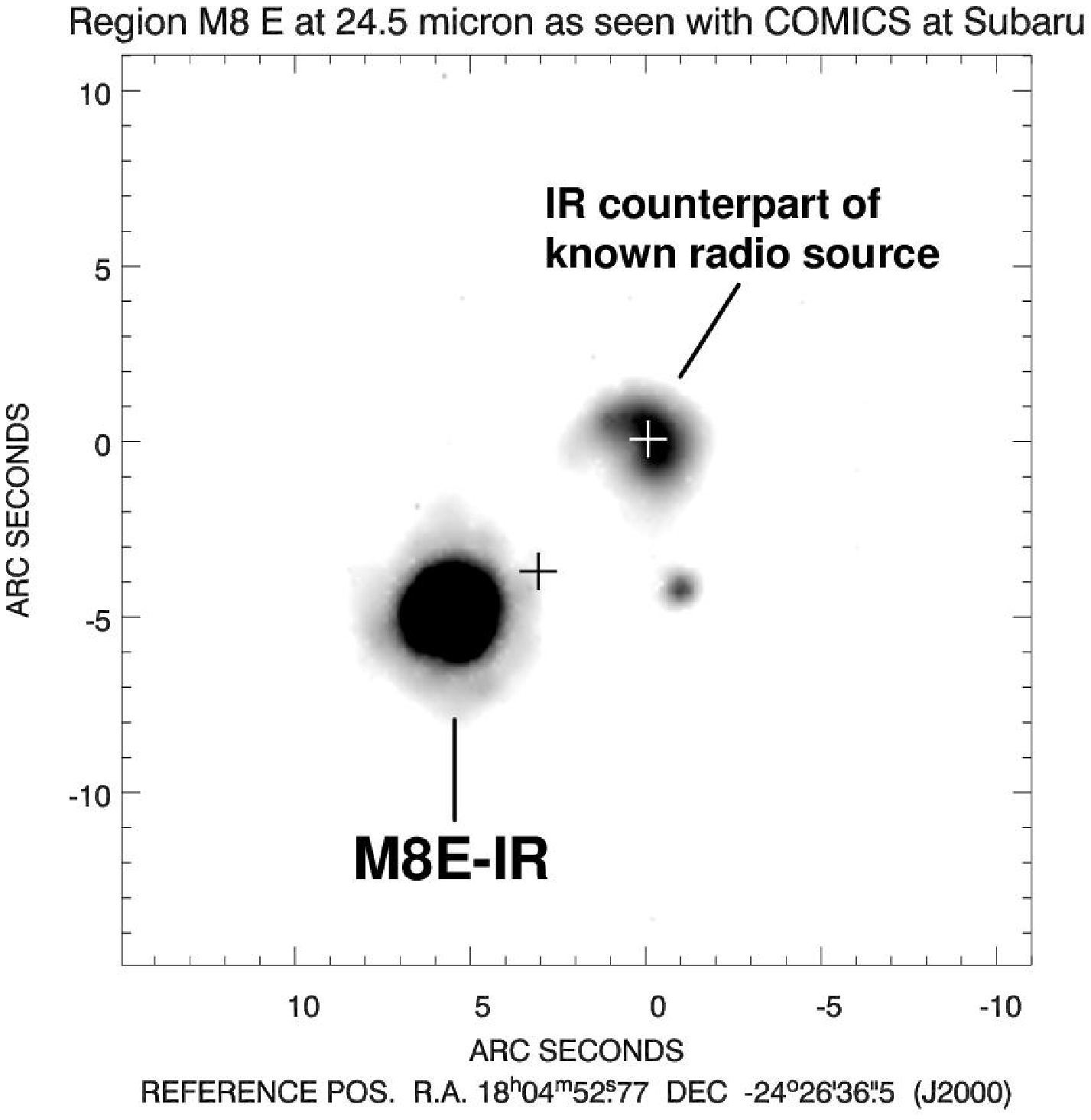}\,\,\,\includegraphics[width=7.75cm]{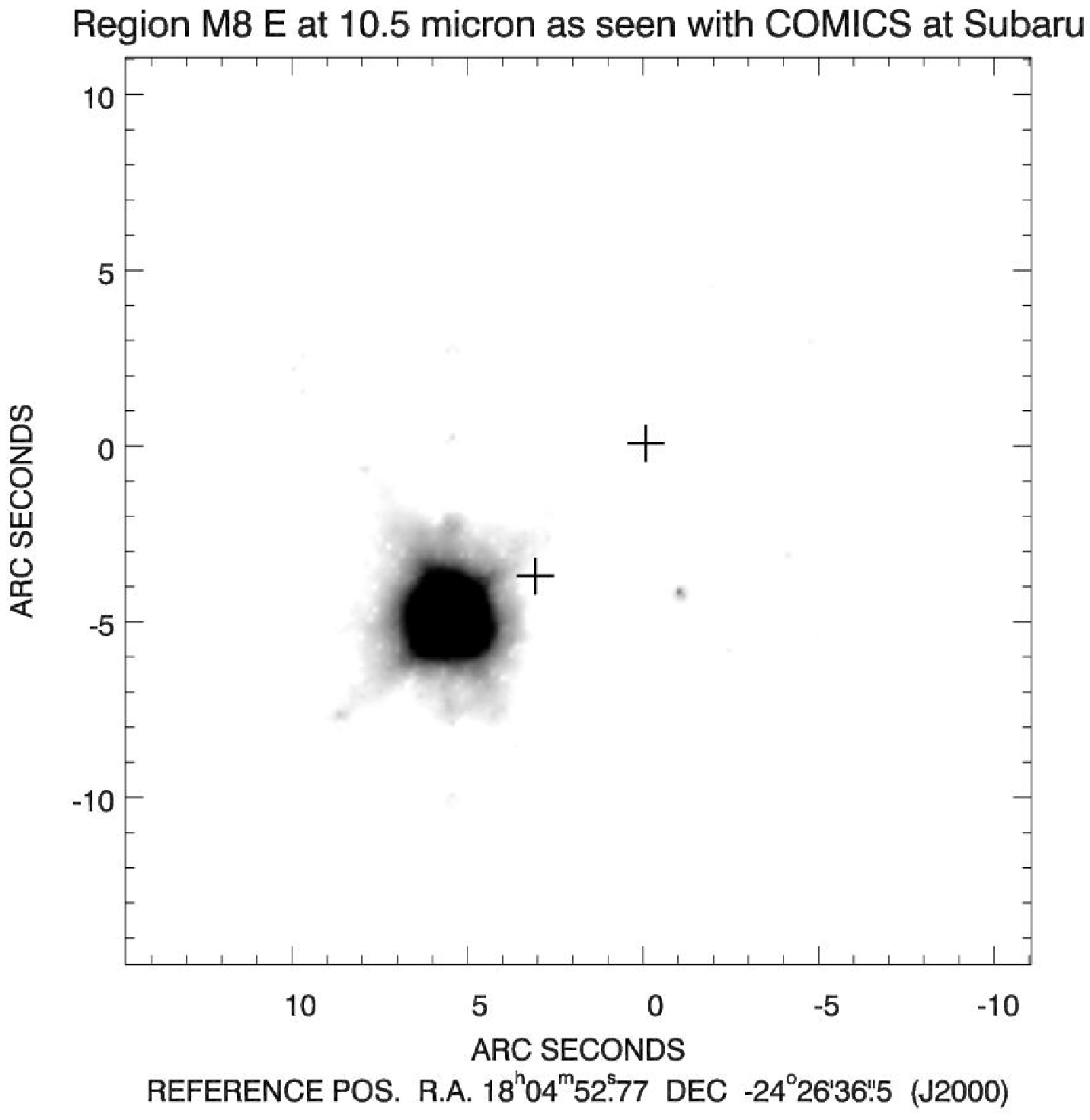}
      \caption{\textbf{Left: } Q-band image of the M8E region at 24.5 $\mu$m with
               COMICS/Subaru. The two previously known objects are annotated.
	       The black plus indicates the position of a 44 GHz methanol maser
	       according to \citet{1999ARep...43..149V}. The white plus
	       marks the position of the cm continuum source 
	       \citep{1998A&A...336..339M}. \textbf{Right: } N-band image of the M8E region at 10.5 $\mu$m with
               COMICS/Subaru. The two plus signs mark the same objects as in the left image. The IR counterpart of the 
	       cm source is not detected, while the third IR source is faintly visible.
              }
         \label{Fig:Subaru}
   \end{figure*}  
   \noindent

\section{Observations}

\subsection{Mid-infrared imaging with COMICS at Subaru}\label{Sec:COMICS-Obs}

The SMOKA data archive \citep{2002ASPC..281..298B} contains 8.2-m Subaru
observations of M8E-IR in the thermal infrared,  
obtained on June 08, 2004 with the camera COMICS 
\citep{2003SPIE.4841..169O}, utilising the 24.5 $\mu$m Q-band filter 
($\lambda_{\rm c}$ = 24.56 $\mu$m, $\Delta \lambda$ = 0.75 $\mu$m)
as well as the 10.5 $\mu$m N-band filter
($\lambda_{\rm c} = 10.48 \, \mu$m, $\Delta \lambda$ = 1.05 $\mu$m). 
For imaging, COMICS employs one 320 $\times$ 240 pixel$^2$ Si:As IBC detector 
from Raytheon with a nominal pixel scale of 0.13 arcseconds. For the 
observations, the whole detector was read out with an elementary exposure 
time of 0.06 s. These exposures were repeated 16 times per chop position. 
Chopping with a frequency of 1 Hz, a nominal throw of 20$''$ and a 
chop angle of 140\degr{} was applied to remove the strong thermal 
background. 
 Since for COMICS/Subaru, the residual pattern after chopping
subtraction is negligible for most purposes \citep[cf.][]{2006ApJ...644L.133F},
additional nodding was not applied during the observations. 
The total integration time (chop-on plus chop-off) is 400 s in Q-band
and 200 s in N band, respectively. The stars \object{HD 148478} (Q-band) and 
HD 169916 (N-band) were observed directly before and after M8E-IR for 
flux calibration.

\subsection{MIDI interferometry at the VLTI}\label{Sec:MIDI-Obs}

   \begin{figure*}
   \centering \hspace*{-0.3cm}
   \includegraphics[width=8.5cm]{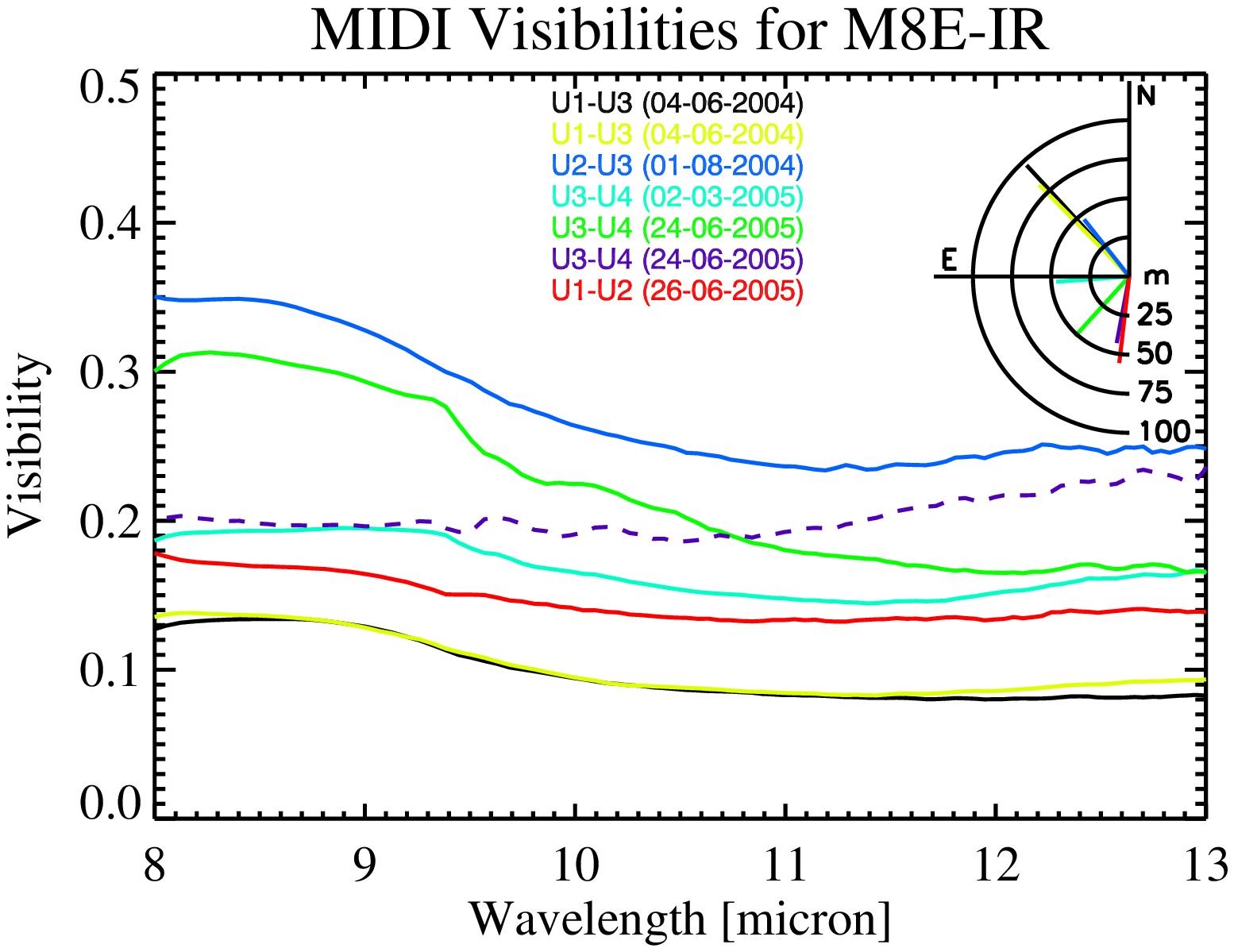}\includegraphics[width=8.5cm]{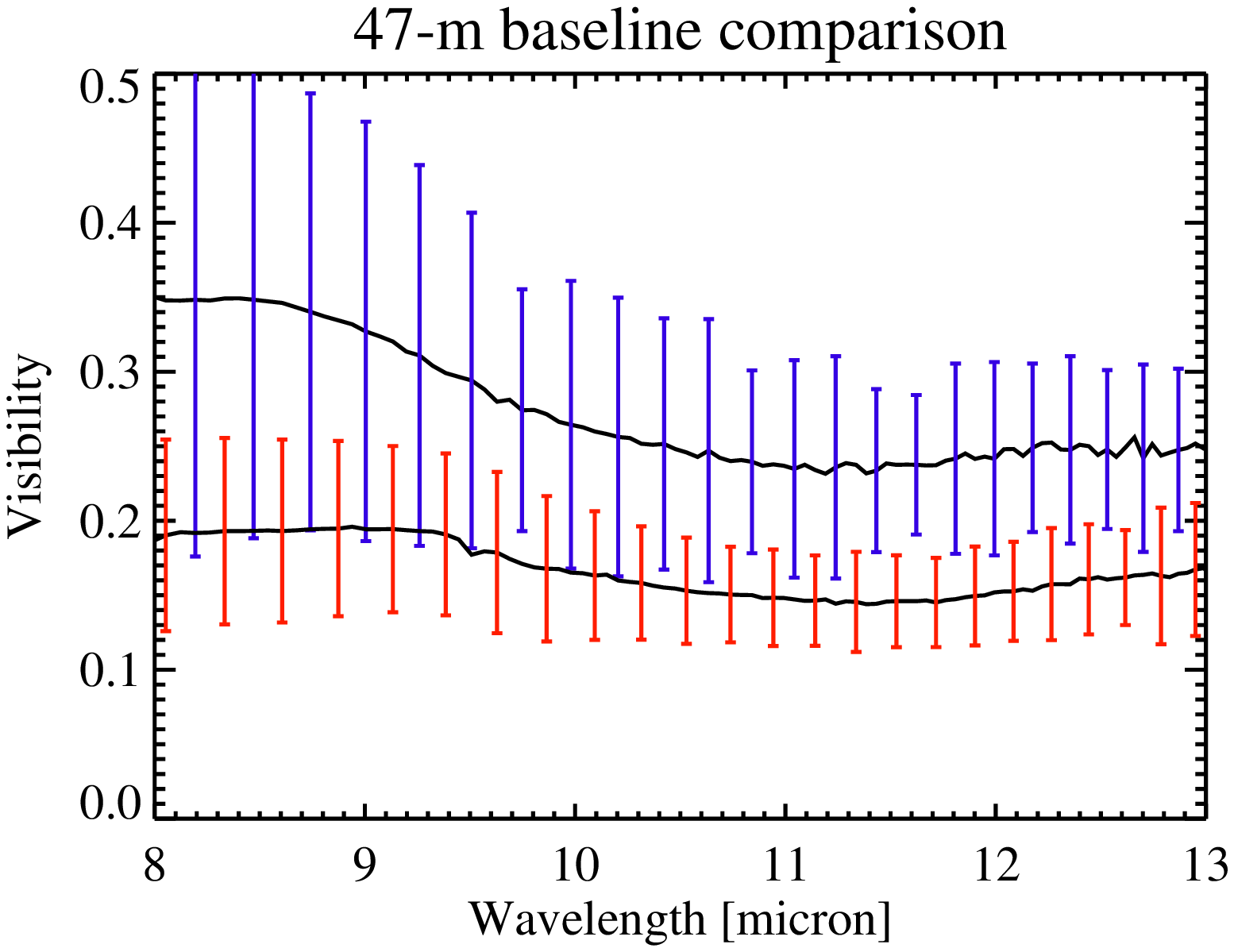}
   \caption{\textbf{Left:} Collection of dispersed MIDI visibilities. The inset
   schematically shows the baseline lengths and orientations 
   (cf.~Table \ref{Tab:MIDI-Log}).  Note that the apparently rising slope of the purple UT3-UT4 visibility 
   curve from 24-06-2005 (dashed line) is probably an artefact of the measurement at very high airmass 
   (see Sec.~\ref{Sec:MIDI-Results}). \textbf{Right:} The two visibility curves with very similar projected
   baseline lengths (46.6 m vs 46.8 m) but strongly differing position angles (+38.4$^\circ$, upper curve vs. 
   $-85.9^\circ$, lower curve). The formal error bars (+/$- 3 \sigma$) are indicated. }
              \label{Fig:VIS}%
    \end{figure*}
    
   \begin{table}[h]
      \caption[]{Log of MIDI observations of M8E-IR.}
         \label{Tab:MIDI-Log}
      \hspace*{-0.25cm}         \begin{tabular}{lcccc}   
            \hline
            \noalign{\smallskip}
    UT \, date \, and \, time  &  B  &  PA   & Telescope & ESO \\
                               & [m] & [deg] &  pair     & Program\\ 
            \noalign{\smallskip}
            \hline
            \noalign{\smallskip}
       2004-06-05 \, 08:07 & 96.8  &    +42.7  &  U1--U3  & 073.C-0175(A)\\
       2004-06-05 \, 09:57 & 82.2  &    +44.5  &  U1--U3  & 073.C-0175(A)\\
       2004-08-01 \, 01:51 & 46.6  &    +38.4  &  U2--U3  & 273.C-5044(A)\\
       2005-03-02 \, 08:58 & 46.8  &  $-$85.9  &  U3--U4  & 074.C-0389(B)\\
       2005-06-24 \, 07:37 & 51.6  &  $-$42.2  &  U3--U4  & 075.C-0755(B)\\
       2005-06-24 \, 09:29 & 43.4  &  $-$10.6  &  U3--U4  & 075.C-0755(B)\\
       2005-06-26 \, 00:26 & 55.7  &  $-$06.6  &  U1--U2  & 075.C-0755(A)\\
            \noalign{\smallskip}
            \hline
         \end{tabular}
   \end{table}

Visibilities in the mid-infrared wavelength range 8--13\,$\mu$m
have been obtained with the instrument MIDI 
\citep{2003SPIE.4838..893L} at the Very Large Telescope Interferometer.
Within the framework of Guaranteed Time Observations for MIDI as well as 
Director's Dicretionary Time, we observed M8E-IR at seven baseline length / 
baseline orientation combinations between June 2004 and June 2005. 
In Table \ref{Tab:MIDI-Log} we list the UT dates
and UT times for the fringe track data, the projected baseline lengths and the
position angles of the projected baselines on the sky (counted from north via 
east on the sky), as well as the telescope configurations used and the observing 
proposal numbers. We refer to \citet{2004A&A...423..537L} for a more detailed 
description of the standard observing procedure for MIDI observations. For all 
our observations, the so-called HighSens mode was used:
during self-fringe tracking, all the incoming thermal infrared 
signal is used for beam combination and fringe tracking, while the 
photometry is subsequently obtained in separate observations. We use the 
MIDI prism as the dispersing element, hence, we finally obtain spectrally 
dispersed visibilities with a spectral resolution of $R  \approx  30$.
\object{HD 169916} was used as the main interferometric and photometric standard 
star and always was observed immediately after M8E-IR. In addition, all calibrator 
measurements of a night were collected to 
create an average interferometric transfer function and to assign error margins
to the measured visibilities. For the August 01, 2004 observations, the conditions
were almost photometric, the airmass of both M8R-IR and HD 169916 was minimal (1.01), 
and the data are hardly affected by atmospheric disturbances (e.g., ozone feature). 
Therefore, we use the dispersed photometry from this measurement to provide the N-band spectrum 
later used in the SED fitting (see Sect.~\ref{Sec:Modelling}).

\section{Results}

\subsection{MIR imaging}
In Fig.~\ref{Fig:Subaru}, we show the 24.5 $\mu$m Subaru/COMICS image  on the left. 
M8E-IR is still the dominating source at this wavelength.
At a nominal resolution of 0\farcs75, the emission remains compact and barely differs from
the profile of the supposedly unresolved calibrator star. In a very recent MIR survey paper by 
\citet{2009A&A...494..157D}, the same Subaru data set is used to analyse the 24.5 $\mu$m intensity profile 
of M8E-IR in more detail and to model the profile and the SED by means of spherically symmetric model 
configurations. These authors come to the same conclusion regarding the compactness of the 
M8E-IR profile.
A second faint point source not yet reported in the literature is visible roughly 6$''$ 
west of it. Furthermore, we clearly detect MIR emission arising from the
neighbouring radio source \citep{1984ApJ...278..170S, 1998A&A...336..339M}.
\citet{1985ApJ...298..328S} had already reported a detection of this source in
N- and Q-band (fluxes only, measured with a 6$''$ diaphragm). For the first time, 
the Subaru MIR imaging spatially resolves this emission.
It is comet-shaped, with the apex directed away from M8E-IR. This morphology
could be an intrinsic property of this UCH{\sc ii} region, or it could be
shaped by the molecular outflow probably arising from M8E-IR 
\citep{1988ApJ...327L..17M, 1992ApJ...386..604M} onto this radio source. 
\citet{2006A&A...459..477D} reported an X-ray source $\approx$ 1\farcs7 southwest of 
M8E-IR (\object{[DFM2004] 845}).  
Accounting for the uncertainty of the X-ray position (2\farcs56), the 
X-ray emission can still be associated with M8E-IR, but not with the 
known radio source or the abovementioned faint MIR point source.\\
At 10.5 $\mu$m we easily detect M8E-IR  (Fig.~\ref{Fig:Subaru}, right). 
The source is very bright, which causes some
additional image artefacts (diffraction spikes in diagonal directions as well as a detector ``drooping'' 
effect, causing multiple fainter copies of the strong source along the north-south direction). 
Hence, we will not further investigate
the shape of M8E-IR in this image. The radio source counterpart is not
detected at 10.5 $\mu$m (rms noise $\approx$ 8 mJy). Considering the COMICS filter 
characteristics (Sect.~\ref{Sec:COMICS-Obs}), combined with the 
detection of the radio counterpart by \citet{1985ApJ...298..328S} in a 
broad N-band filter (0.97 Jy), this implies a very strong and broad 9.7 $\mu$m 
silicate absorption feature  or time variability. The third source 
in the 24.5 $\mu$m image is detected at a $4 \sigma$ level at 10.5 $\mu$m.

\subsection{MIR interferometry}\label{Sec:MIDI-Results}
We reduced the interferometric data with the MIA+EWS package, version 1.5,
developed at the MPIA Heidelberg and the University of Leiden. The resulting 
visibility curves are collected in Fig.~\ref{Fig:VIS}, left. 
The object is clearly resolved in all configurations with visibilities between 0.09--0.35. 
If we assume a Gaussian intensity distribution\footnote{The relation between the visibility 
and the intensity full width half maximum is given by 
$V(u)$ = exp($- \pi^2/(4 \, $ln$ 2) \ \  u^2 \,  FWHM^2$) in this case, where the 
spatial frequency is $u = \lambda/B$, with $B$ being the modulus of the projected baseline
length on the sky.}
of the source, the visibilities indicate an intensity FWHM of $\approx$ 20--25 mas (8.5 $\mu$m) and 
32--38 mas (12.0 $\mu$m) which is in rough agreement with the extension of the small 
component of \cite{1985ApJ...298..328S}. We note that these visibilities, although
not reaching the relatively high levels of most Herbig Ae/Be stars
\citep[e.g.,][]{2004A&A...423..537L}, are qualitatively different
from the very low visibilities (0.01--0.05) found for several of the
other objects in our sample as well as recently reported for two other massive
YSOs \citep{2007ApJ...671L.169D,2008ASPC..387..444V}. Still, to learn more about 
the potentially more complicated intensity structure, a simple Gaussian is not a 
sufficient ansatz, especially if the object is strongly resolved by the interferometer.
Further modelling is necessary for interpretation.\\
Can we infer the geometry of the source from our visibility curves alone? Most visibilities 
have been taken at different baseline length / position angle combinations. Still, there are two 
measurements with almost the same projected baseline length (46.6 m vs 46.8 m), but position angles
differing by 55.7$^\circ$ (cf. Table \ref{Tab:MIDI-Log}). A spherically symmetric intensity distribution
will result in visibilities not being a function of position angle. These two measurements, however, 
show different visibility levels (Fig.~\ref{Fig:VIS}, right). But when considering the 3-$\sigma$ error bars,
this difference is not fully conclusive, since the error bars, at least in the 8 to 11 $\mu$m range, still
overlap slightly. At longer wavelengths, the two error intervals are almost distinct. While we cannot draw
strong conclusions from these findings, this is a first hint that the mid-infrared intensity distribution
of M8E-IR is indeed not fully spherically symmetric. \\				
Visibilities can be considered as the ratio of correlated flux to total flux. The N band total flux spectrum of 
M8E-IR shows a moderately deep silicate absorption spectrum (see inset of Fig.~\ref{Fig:SED}). Most of the 
curves displayed in Fig.~\ref{Fig:VIS} (left) show a drop in the visibility level over the silicate
feature. However, the visibility curve corresponding to the 2005-06-24 09:29 UT measurement does not show this 
behaviour, but instead rises monotonically from 8 to 13 $\mu$m. We think that this is an artefact of the
observation. While all the other measurements were taken at intermediate elevations (airmass $<$ 1.73), the
data set in question was obtained at a very low elevation (airmass 2.41) in order to get this (u,v) point, 
corresponding to a relatively short projected baseline length. At such a low elevation, the performance of MIDI 
is certainly affected. Since the adaptive optics system cannot be locked on M8E-IR itself, but only on a 
faint optical star more than 40$''$ away, the beam stability and therefore the beam overlap on the MIDI detector 
may be compromised. Since the PSF is smaller at 8 micron than at 13 micron, a smaller beam overlap will lead to 
apparent losses in correlated flux preferentially on the short-wave side of the spectrum. We have critically checked 
the overlap for this measurement and indeed find that the two beams are more separated than for other observations 
of M8E-IR as well as for the calibrators. By artificially restricting the overlap mask to a smaller central area,
the slope of the visibility curve can be flattened with (calibrated) visibility values at 8 micron slightly higher 
than at 13 micron. Hence, the low 8 micron correlated flux of the indicated visibility curve is an artefact of the 
measurement process. Therefore, we will not further interpret the shape of this particular visibility curve.

\subsection{Modelling}\label{Sec:Modelling}

   \begin{figure}
   \includegraphics[width=\linewidth]{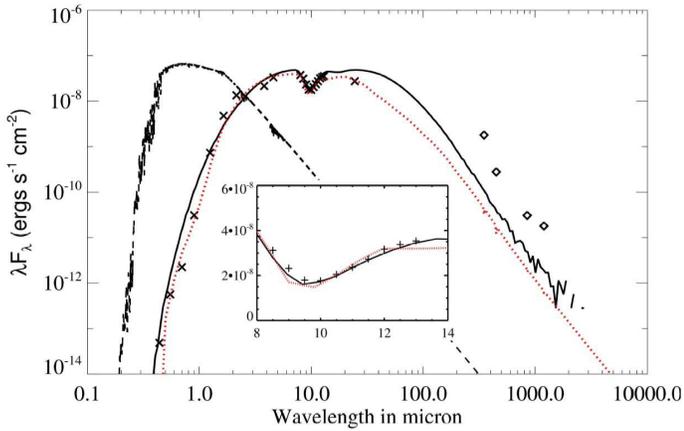}
      \caption{SED of M8E-IR. Crosses = observed fluxes,
   diamonds = upper limits. The solid line gives
   the best-fitting radiative transfer model  3007526 from the grid of models by
   \citet{2007ApJS..169..328R} (effective synthetic aperture $\sim 2''$).
   The dashed  black curve shows the SED of the bloated central star used in that model.
    The red dotted line shows the spherical model with a hot star (also mentioned in 
   Sec.~\ref{Sec:Modelling}) that fits the SED but not the
   visibilities (cf.~Fig.~\ref{Fig:RT-Comp}).
   The inset shows a zoom into the 10\,$\mu$m region using a synthetic aperture 
   $<1''$, appropriate for the underlying MIDI spectroscopy data (shown as plus signs). 
   For legibility, the symbol sizes are similar to or larger than the real error bars.}
         \label{Fig:SED}
   \end{figure}
\begin{figure}
   \includegraphics[width=\linewidth]{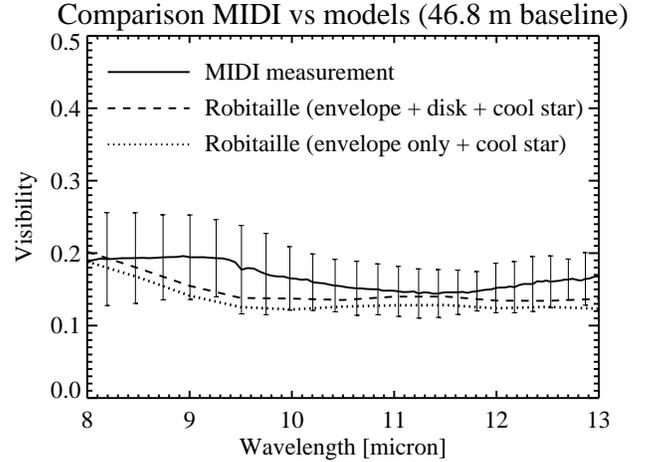}
      \caption{Spectral shape comparison of the visibilities for the 46.8-m baseline as an
      example. The MIDI measurement plus 3-$\sigma$ error bars is given as a solid line. Furthermore, 
      radiative transfer synthetic images between 8 and 13 micron  have been produced for the best 
      envelope + disk model (dashed line) and the best envelope-only model (dotted line). Cuts through 
      their spatial frequency spectrum corresponding to the 46.8-m baseline are concatenated to display 
      the synthetic spectral visibility shape of these models.
}\label{Fig:46.8m-Comp}
   \end{figure}

\begin{figure}
   \centering
   \includegraphics[width=9cm]{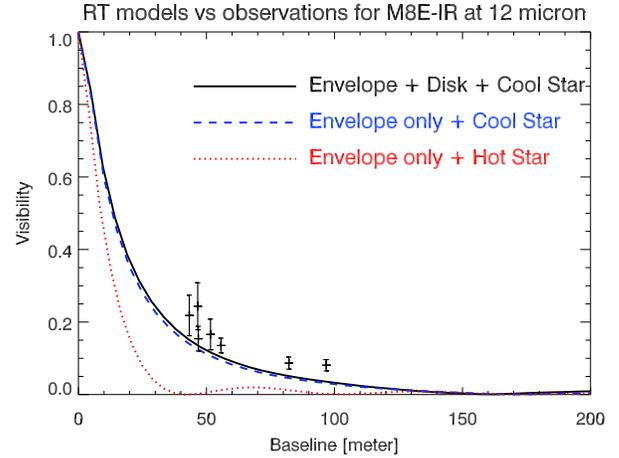}
      \caption{
      Cuts through the 12 $\mu$m spatial frequency spectrum (in units of the 
      interferometric baseline length) of 3 models 
      well fitting the SED. The black solid line refers to a model with a small
      circumstellar disk, flattened envelope and cool bloated central object 
       (Table \ref{Tab:Parameters}). The blue dashed 
      line denotes a model with a flattened envelope and cool bloated central object without
      a disk; the intensity distribution is almost identical to the
      first model. Due to the low inclination of these 2D models (18$^\circ$),
      the central intensity distribution shows only small deviations from radial
      symmetry. Hence, differences in the model visibilities with varying position 
      angle remain small ($<10$\%), and the cuts can be 
      adopted as representative. The classical approach (hot central star
      + spherical envelope, red dotted line), however, gives far too low visibilities 
      compared to the MIDI data (plotted as plus signes including 
      the formal 3$\sigma$ error bars).}\label{Fig:RT-Comp}
   \end{figure}

  
We apply self-consistent continuum radiative transfer modelling to M8E-IR in 
order to produce synthetic MIR intensity maps and to compare their spatial 
frequency spectrum with the observed visibilities. Self-consistency here
means that the dust temperature is calculated from the condition of (radiative) 
energy conservation. Hence, the luminosity sources in the system
(central object + accretion) initially heat up the dust until a radiative
equilibrium is reached, the temperature (for the given dust density structure)
converges, and the total luminosity is recovered in the emerging SED 
\citep{2001ApJ...554..615B, 2003ApJ...591.1049W, 1997A&A...318..879M}. 
We are mainly concerned
with the question of which spatial distribution of the circumstellar material can 
account for both the SED and the visibilities of M8E-IR. \\
For SED fitting, we use the M8E-IR photometric data collected in 
\citet{2002ApJS..143..469M} plus new 1.2 mm data from \citet{2006A&A...447..221B}. 
We stress that no (sub-)millimeter interferometry 
on M8E-IR is reported in the literature which could spatially disentangle 
the flux contributions from M8E-IR and the radio source 8$''$ away. Hence, we 
consider the fluxes for $\lambda > 24.5 \, \mu$m as upper limits in 
our modelling. Furthermore, we include new optical photometry in the B, V, and 
I filters reported by \citet[][their object \object{Cl* NGC 6530 WFI 13458}]{2005A&A...430..941P} 
as well as our new 24.5 $\mu$m photometry (220 $\pm$ 20 Jy). The 8--13 $\mu$m total
flux spectrum taken during the MIDI measurements 
(Sect.~\ref{Sec:MIDI-Obs}) was used to further curtail the number of viable 
models. \\
   \begin{table}[t]	
      \caption[]{Parameters of the best-fitting Robitaille model 3007526.}
         \label{Tab:Parameters} 
      \hspace*{-0.25cm}         \begin{tabular}{lrlr}   
            \hline
            \noalign{\smallskip}
    Parameter  &  Value  &  Parameter  & Value \\
            \noalign{\smallskip}
            \hline
            \noalign{\smallskip}
       Effect.~temperature        & 4,740 K         &  Stellar radius            &  125 R$_\odot$  \\
       Stellar mass               & 13.5 M$_\odot$  &  Disk mass                 &  0.71 M$_\odot$ \\
       Outer disk radius          & 16 AU           &  Density exp.$^{(1)}$      &  2.05  \\
       Scale height exp.$^{(1)}$  & 1.05            &  Viscosity $\alpha$ par.$^{(1)}$  &  1.3E-2    \\
       Envelope radius            & 6.8E+4 AU       &  Mass infall rate$^{(2)}$  &  7E-5 M$_\odot$/yr  \\
  $A_{\rm V}^{\rm tot}$ along LOS & 33 mag          &  Inclination angle         &  18$^\circ$  \\
       Cavity cone angle  	  & 3.3$^\circ$	    &				 &  \\
            \noalign{\smallskip}
            \hline
         \end{tabular}
	 $^{(1)}$ for the disk   \hspace{0.15cm} $^{(2)}$ for the envelope  
	 \hspace{0.45cm} Dust model: \citet{1994ApJ...422..164K}
   \end{table}
We used the SED online fitting tool of \citet{2007ApJS..169..328R}  
that can in principle utilise models including an
envelope plus circumstellar disk. We refer to this publication for details
on the setup of these models.  To produce the synthetic MIR intensity maps corresponding to the
Robitaille models, we use the underlying Monte Carlo radiative transfer code of \citet{2003ApJ...591.1049W}. The SED 
fitting tool delivers the full parameter set for the best-fitting models which can be directly included in the 
Whitney code. The resulting images are then Fourier-transformed, and cuts through this spatial frequency spectrum of the
images give the synthetic visibilities. \\
Intriguingly, we do not find a unique solution from this SED fitting. 
The  formally best SED fit points to a model featuring a very compact circumstellar disk ($<$ 20 AU), a larger
surrounding envelope with small bipolar cavities and a bloated cooler
central object  (cf.~Table~\ref{Tab:Parameters}) In Fig.~\ref{Fig:SED}, 
we show this best SED fit to M8E-IR by the Ro\-bi\-tail\-le models 
 as black continuous line. 
Due to the linkage of the Ro\-bi\-tail\-le grid to evolutionary tracks, certain 
size parameters cannot be independently chosen by the fitter \citep{2008ASPC..387..290R}. 
This especially affects the possible disk size which is not well constrained in our case. 
The data would still allow for a somewhat larger disk.  We have repeated the calculation of the
intensity maps with all model parameters fixed to the values of the model best fitting the SED, but have
used larger disk sizes of 50, 100, 250, and 500 AU. For disk radii $\ga 100$ AU, the resulting 
MIR visibilities eventually drop. Hence, the difference to the values measured with MIDI increases,
which makes such solutions less likely. \\ 
Among the well-fitting  Robitaille SED models are also configurations 
without a  compact disk (axisymmetric flattened envelope + outflow cavities only), 
which nevertheless give almost identical elevated visibilities  (see Fig.~\ref{Fig:RT-Comp}). 
This suggests that in the case of M8E IR, the choice of the central object actually governs
the resulting visibility levels (see below). \\
 We have also compared the spectral shape of the visibility curves with the models. For the comparison,
we have computed synthetic images based on the parameters from the SED fits at eleven wavelengths between 8 
and 13 micron. Their Fourier transform is then probed at spatial frequencies according to the projected
baseline lengths of the MIDI measurements. In general, the Robitaille models predict somewhat lower visibilities 
than indicated by the measurements. The general shape of all the synthetic visibility curves is similar. 
Therefore, we show as one example the 46.8-m baseline comparison in Fig.~\ref{Fig:46.8m-Comp}. We compare 
the best envelope + disk configuration (Robitaille model 3007526, see Table \ref{Tab:Parameters})
and the best envelope-only configuration (Robitaille model 3004120), both having a cool central object (see also
Fig.~\ref{Fig:RT-Comp}), to the MIDI measurement. The model visibility curves show a clear decrease in the visibility level toward the
amorphous silicate feature. The modelled shape would arise naturally in the circumstellar disk context when the correlated flux
comes mainly from an inner, optically thicker region with a somewhat suppressed silicate emission feature, while the total 
flux is dominated by more optically thin emission from the wider disk surface \citep{2005A&A...441..563V}. However, as 
seen in Fig.~\ref{Fig:46.8m-Comp}, the envelope-only model also gives the same behaviour. 
Hence, based on the modelled visibilities, it is not possible to distinguish between the two cases for our special setup.
This is understandable considering the compactness of the disk in question, which would not be resolved even by our longest 
baseline measurement. The envelope by far dominates the mid-infrared emission of the whole object. 
The measured visibilities also show a level decrease. However, it is not centered around 9.7 micron, but occurs at longer 
wavelengths. Currently, we do not understand this shift. It is not clear if it
can be fully explained by differing dust properties in M8E-IR and assumed for the radiative transfer simulations
\citep[opacities and grain size model from][]{1993ApJ...402..441L, 1994ApJ...422..164K}. To further analyse and reproduce 
the spectral shape details of the
visibility curves is beyond the scope of this paper.\\
We explicitly mention the low inclination angle of $<30^\circ$ inherent in all 
our  Robitaille models fitting the SED data. This is in contrast to the disk
hypothesis of \citet{1985ApJ...298..328S} that featured a very large
inclination. However, Simon et al.~saw a more or less
symmetric intensity distribution in their 10 $\mu$m lunar occultation data;
hints for deviations came mainly from lunar occultations recorded at 3.8 $\mu$m.
At this shorter wavelength, configurations with disk + envelope including outflow
cones are naturally more structured, and scattered light in the inner regions 
of these cones can still contribute, in particular if grains larger than the
typical 0.1 $\mu$m dust particles are involved. \\
\noindent
   
\subsubsection{Comparison with canonical 1D models}
As mentioned above, the Robitaille models are not optimally
sampled in the parameter ranges appropriate for massive YSOs. Formally, the Robitaille
models cover a sufficiently large parameter space regarding the 14 important parameters. In particular, 
hot, non-bloated central stars are included, up to effective temperatures of $>50,000$ K 
\citep[see Fig.~3 in][]{2006ApJS..167..256R}. However, not all parameter combinations are available for
the fitting since it is not possible to explore the huge parameter space in its entirety in a completely
unbiased manner when constructing the model grid. Hence, the set of models available for the SED fitting does 
not form a complete ``grid'' in a technical sense, since parameters are randomly 
sampled within ranges \citep[cf.][especially the sections 2.1.4 and 2.2.1]{2006ApJS..167..256R}. Since we 
have found quite peculiar models from our fitting attempt (see above), we felt compelled to do a 
cross-check of our results.
We could have missed simpler models among the top-rank fitting results that nevertheless agree with 
the observed SED. Traditionally, such BN-type objects
have been modelled as a spherical dust shell surrounding a hot, non-bloated central
star \citep[e.g.,][]{1991A&A...252..801G, 2009A&A...494..157D}.  To check if even such canonical 
configurations can account for the SED and the visibilities, we employed 
well-tested models, based on the code used by \citet{1997A&A...318..879M, 1999ApJ...519..257M}.
Unlike the Whitney code, the Men'shchikov program is a grid-based radiative transfer code. It comes with
its own ray-tracing module to produce high-definition intensity maps. Here, we can
tune the parameters without being restricted by a predefined parameter grid, and the fine-tuning
process in the case of pure 1D models can be much faster.
With a spherically symmetric envelope  (density power law $\sim r^{-1.7}$, 
$M_{\rm env}=40 \ {\rm M_\odot}$) and a 30,000 K central star  with 3.7 R$_{\odot}$,
it is possible to fit the SED of M8E-IR  well (Fig.~\ref{Fig:SED}, see also 
\citet{2009A&A...494..157D}\footnote{These authors find slightly different model parameters, 
because they try to find a balance between fitting the 24.5 $\mu$m intensity profile and properly fitting 
the SED. Thereby, they take some of the existing single-dish (sub-)millimeter points as face value. 
There are indications, however, that at long wavelengths the neighbouring UCH{\sc ii} region 
(Fig.~\ref{Fig:Subaru}, left) dominates the emission of the region. Therefore, these long wavelengths 
fluxes should be considered as strict upper limits.}). 
Still, compared to the measured visibilities, these models result in far too 
low visibilities ($<$ 0.05) for M8E-IR in the baseline range 30--60 m 
over the whole 8--13 $\mu$m range, even when we add a population of 
very large grains (200\,$\mu$m) to the central shells to raise the 
visibilities. The error margins of the MIDI visibilities 
(on average 10\%) do not account for such large differences.
Independently, we found that the inclusion of a cold ($\sim 2,000$~K) 
central object will sufficiently raise the visibilities also in the 
Men'shchikov models.
The circumstances are illustrated in Fig.~\ref{Fig:RT-Comp} where the (u,v)-spectra 
of the 12 $\mu$m synthetic images based on  3 different models are compared. 
 At such a longer wavelength outside the silicate absorption feature, the influence
of different dust models on the results should be small. Furthermore, potential observational
biases due to decreased beam overlap during the MIDI measurements are smaller at longer 
wavelengths (see  Sec.~\ref{Sec:MIDI-Results}).
Although the (u,v)-spectra of the  cool star models still show somewhat lower 
visibilities than the measured ones  (the visibilities are not included in the 
fitting loop), obviously they are qualitatively different 
from the  hot star model. \\

\subsubsection{Feasibility of a cool central object}
Interestingly, the best-fitting models for M8E-IR in the Ro\-bi\-tail\-le model grid 
show central stars of 10--15 M$_\odot$ that are strongly bloated and, 
therefore, have relatively low effective temperatures  ($<$ 5000 K). 
Such solutions can occur since the Ro\-bi\-tail\-le grid also comprises the 
full range of canonical pre-main sequence evolutionary tracks from the Geneva 
group\footnote{http://obswww.unige.ch/$\sim$mowlavi/evol/stev\_database.html} 
as possible parameterisation of the central objects. Although M8E-IR cannot 
straightforwardly be identified with corresponding very early evolutionary
stages, the tendency for a bloated central star in the case of M8E-IR may be 
valid. As found already by \citet{1977A&A....54..539K}, accretion with high 
rates onto main sequence stars can temporarily puff up such stars.
Furthermore, \citet{2008ASPC..387..255H,2009ApJ...691..823H} and 
\citet{2008ASPC..387..189Y} have recently computed the pre-main-sequence 
evolution of stars as a function of the accretion rate onto the forming star. 
They find that for accretion rates reaching 10$^{-3} {\rm M}_\odot$/yr, 
the protostellar radius can temporarily increase to $> 100 \,{\rm R}_\odot$, 
in accordance with our model fitting.
Note that \citet{1988ApJ...327L..17M} revealed high-velocity molecular 
outflows from M8E-IR based on M-band CO absorption 
spectra and speculated on recent ($<120$ yr) FU Ori-type outbursts for this 
object. If these multiple outflow components really trace recent strong 
accretion events, the central star indeed could have been affected. 
Furthermore, M8E-IR is not detected by cm observations with medium 
sensitivity \citep{1984ApJ...278..170S, 1998A&A...336..339M}. This could be 
explained by large accretion rates still quenching a forming hypercompact 
H{\sc ii} region \citep[e.g.,][]{1995RMxAC...1..137W}. Similarly, 
a bloated central star with $T_{\rm eff} \ll 10000$ K also gives a natural 
explanation of these findings.
   
\section{Conclusions}
We have observed the BN-type massive young stellar object M8E-IR with the
MIDI MIR interferometer at the VLTI. We find substructures with MIR sizes 
around 30 mas. 
 The measured elevated visibilities indicate that the usual approach for BN-type
objects (a spherically symmetric envelope with a hot central star) 
fails in the case of M8E-IR. Most interestingly, our data are consistent with  
M8E-IR harbouring a 10--15 M$_\odot$ central star that has been 
bloated  ($R > 100 \, {\rm R}_\odot$) by recent strong accretion events. 
However, with the present data we 
cannot clearly infer the existence of a circumstellar disk since disk-envelope and 
envelope-only models result in very similar SEDs and visibilities.
If a disk exists it cannot be large ($\la$ 100 AU).
Until (sub-)millimeter observations with sufficient spatial
resolution constrain the long-wavelength emission, disks with radii
from 15--100 AU seem probable.
Both aspects, the central star bloating and the existence of a circumstellar disk,
should be the topic of future investigations. For instance, by utilising 
high-resolution IR spectroscopy one can test whether hydrogen recombination lines show 
imprints of disk rotation along the guidelines of the disk-wind simulations by 
\citet{2005MNRAS.363..615S}.


\begin{acknowledgements}
Based in part on Subaru Telescope data and obtained from the 
SMOKA data archive, which is operated by the Astronomy Data Center at
NAOJ. We thank Thomas Robitaille, 
Barbara Whitney, and Melvin Hoare for discussions on the subject.
Valuable comments from the referee are appreciated.
\end{acknowledgements}

\bibliographystyle{aa}
\bibliography{Linz}

\newpage



\end{document}